\documentclass[prl,aps,twocolumn,reprint,superscriptaddress]{revtex4-1}
\usepackage{graphicx}
\usepackage{braket}
\usepackage{bm}
\usepackage{amsmath}
\usepackage{amsfonts}
\usepackage{graphicx}
\usepackage{color}

\newcommand{\len}[0]{\mathrm{L}}

\newcommand{\ham}[0]{\mathcal{H}}

\begin{document}

\title{Light-matter decoupling and $\mathbf{A}^2$ term detection in superconducting circuits}

\author{J.\ J. \surname{Garc\'ia-Ripoll}}
\email{jj.garcia.ripoll@csic.es}
\affiliation{Instituto de F\'isica Fundamental IFF-CSIC, Calle Serrano 113b, Madrid E-28006, Spain}
\author{B. Peropadre}
\affiliation{Department of Chemistry and Chemical Biology, Harvard University, Cambridge MA, 02138}
\author{S. \surname{De Liberato}}
\affiliation{School of Physics and Astronomy, University of Southampton, Southampton, SO17 1BJ, United Kingdom}

\begin{abstract}
We study the spontaneous emission of a qubit interacting with a one-dimensional waveguide through a realistic minimal-coupling interaction. We show that the diamagnetic term $\mathbf{A}^2$ leads to an effective decoupling of a single qubit from the electromagnetic field. This effects is observable at any range of qubit-photon couplings. For this we study a setup consisting of a transmon that is suspended over a transmission line. We prove that the relative strength of the $\mathbf{A}^2$ term is controlled with the qubit-line separation and show that, as a consequence, the spontaneous emission rate of the suspended transmon onto the line can increase with such separation, instead of decreasing.
\end{abstract}

\maketitle

When the vacuum Rabi frequency, $\Omega$, of an electromagnetic mode is much smaller than the bare frequency of the excitation to whom it couples, $\omega$, the simple Jaynes-Cumming or Tavis-Cumming models capture the main features of light-matter interaction and cavity QED\ \cite{Haroche,Kavokin}. However, already for a normalised coupling $\frac{\Omega}{\omega}>0.1$, the Rotating Wave Approximation (RWA) that justifies those solvable models fails\ \cite{Ciuti05}. In this ultrastrong coupling (USC) regime, light-matter interaction must be described beyond RWA, using the Rabi\ \cite{Braak11} and Hopfield-Bogoliubov\ \cite{Hopfield58} models, that correctly describe the ground state squeezing and asymmetric splitting \cite{Ciuti06,Gunter09,Auer12,Ridolfo13}. The USC regime, observed for the first time only few years ago\ \cite{Anappara09}, has now been achieved in many solid-state cavity quantum electrodynamics setups\  \cite{Niemczyk10,Todorov10,Muravev11,Schwartz11,Geiser12,Scalari12,Gubbin14}, with an actual coupling record of  $\frac{\Omega}{\omega}=0.87$ \cite{Maissen14}. When the normalised coupling becomes of the order one, also the aforementioned non-RWA models fail. In this regime, named deep strong coupling (DSC)\ \cite{Casanova10}, the localised dipolar interaction dominates and a real-space description with many excited photonic modes becomes essential.

Our understanding of such a deep non-perturbative regime is still incomplete\ \cite{Bamba13,DeLiberato14b,Bamba14,Bamba14c}. A first, recent counter-intuitive result is that light and matter eventually decouple in the DSC regime: the spontaneous emission rate of the system dramatically decreases, instead of increasing, with the coupling strength\ \cite{DeLiberato14}. This decoupling is associated with the presence of the diamagnetic term $\mathbf{A}^2$, that expels modes away from the emitter. Still, the decoupling has been rigorously proved only for linear systems ---a perfect planar metallic cavity coupled to a 2D sheet of dipoles---, and the link of the decoupling effect with the $\mathbf{A}^2$ term remains a hypothesis. Indeed, without diamagnetic term, the model in Ref.\ \cite{DeLiberato14} becomes unstable and undergoes a superradiant phase transition\ \cite{Birula79,Lambert04,Bamba14b}, impeaching a comparison of DSC physics with and without $\mathbf{A}^2$.

\begin{figure}
\includegraphics[width=0.75\linewidth]{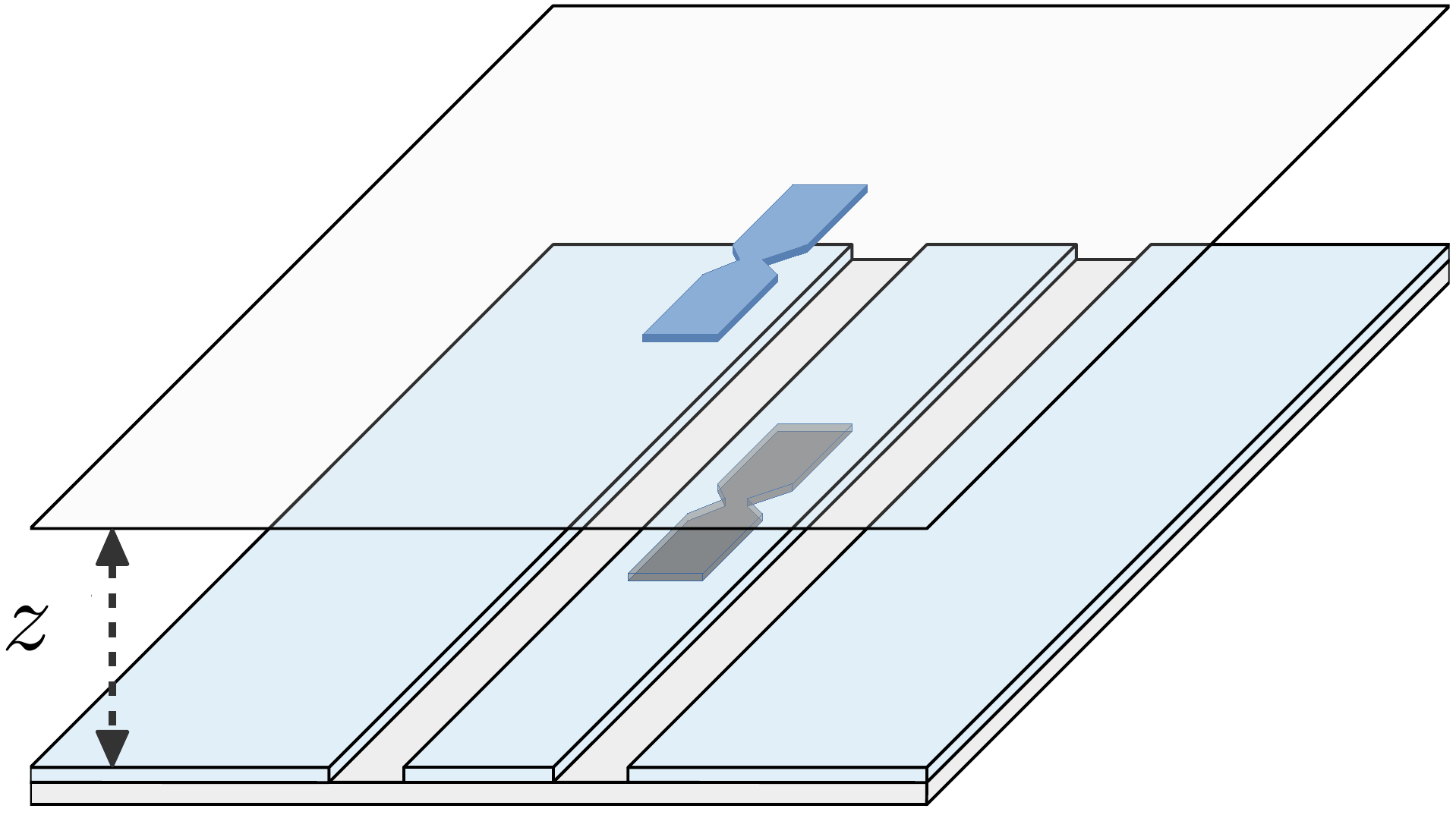}
\caption{The spontaneous emission of a  superconducting qubit, such as a transmon, suspended on top of a transmission line depends non-monotonically on the separation from the line, $z$, because of the influence of the $\mathbf{A}^2$.}
\label{fig:setup}
\end{figure}

The first major result of this Letter is to prove that this decoupling effect happens \textit{already at the single dipole} or qubit limit, at all levels of interaction strength. We arrive at this result by studying the nonlinear interaction between a two-level system and a one-dimensional waveguide, modeled by the Ohmic spin boson model\ \cite{Peropadre13,Leggett87}, and proving that the spontaneous emission rate of the two-level system decreases due to the $\mathbf{A}^2$ term.

The second major result is the possibility of measuring the $\mathbf{A}^2$ term and its effect, using a superconducting transmon qubit\ \cite{Koch07,Schreier08} that is suspended on top of a microwave guide [cf.\ Fig.\ \ref{fig:setup}]. This setup, which profits from the ever improving coherence properties\ \cite{Devoret13,Nori11} and strong interaction\ \cite{Wallraff04,Niemczyk10,Forn-Diaz10} of superconducting circuits, was introduced in Ref.\ \cite{Shanks13} as an ultrasensitive scanning probe that can, among other things, analyze locally complex quantum simulators\ \cite{Houck12}. In this work we prove that the same setup allows a control of the relative strength of the diamagnetic term through the separation between the qubit and photon planes. The striking consequence of this control is that the coupling between the qubit and the line first increases and then decreases as the qubit is moved away from the line. Moreover, this non-monotonic behavior is observable with \textit{existing} transmon technology and does not need of DSC/USC regimes. Finally, the same setup can be used to explore other effects of the $\mathbf{A}^2$ term, including many-qubit phase transitions\ \cite{Nataf10b}.

\begin{figure*}[t]
  \centering
  \includegraphics[width=0.33\linewidth]{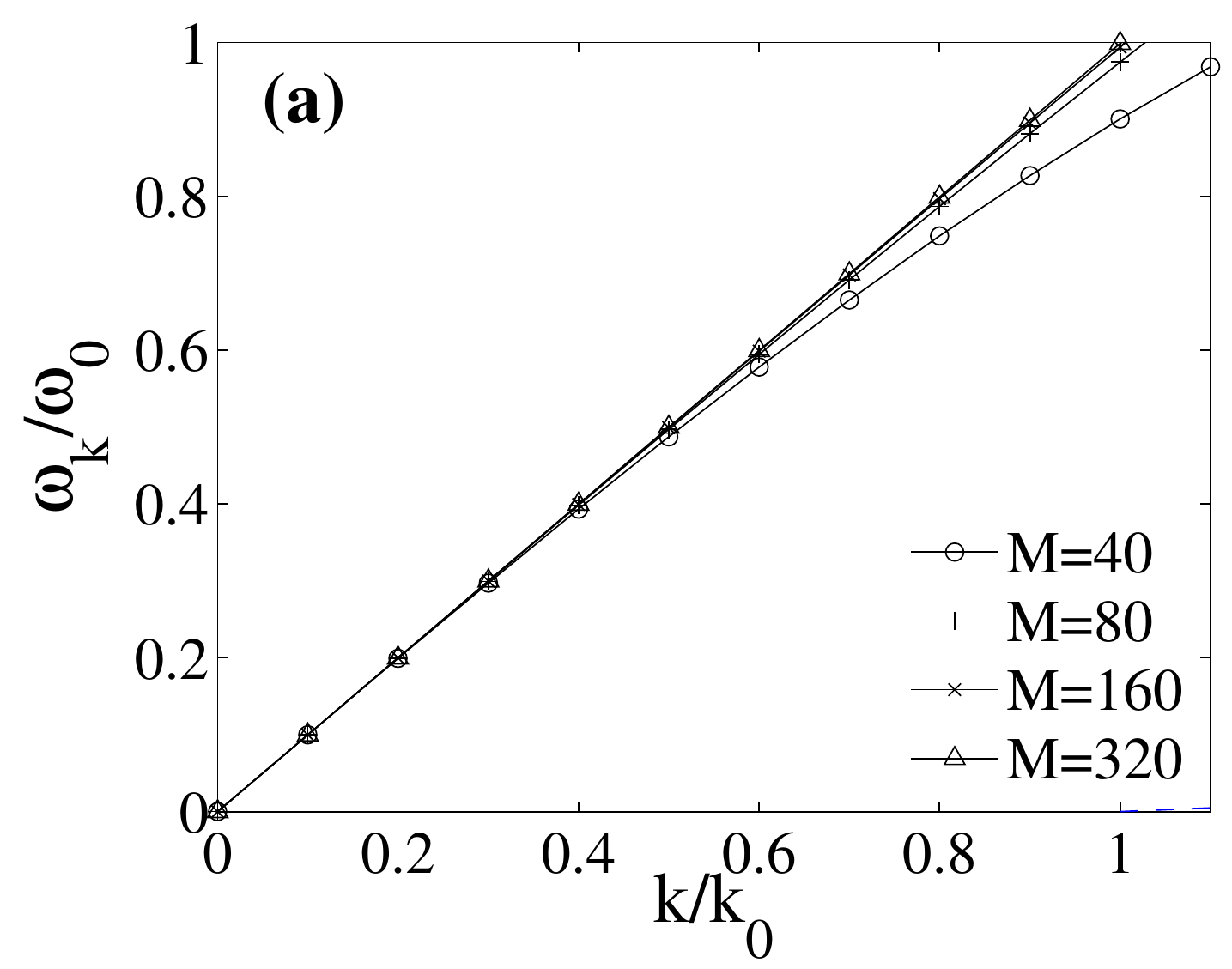}%
  \includegraphics[width=0.33\linewidth]{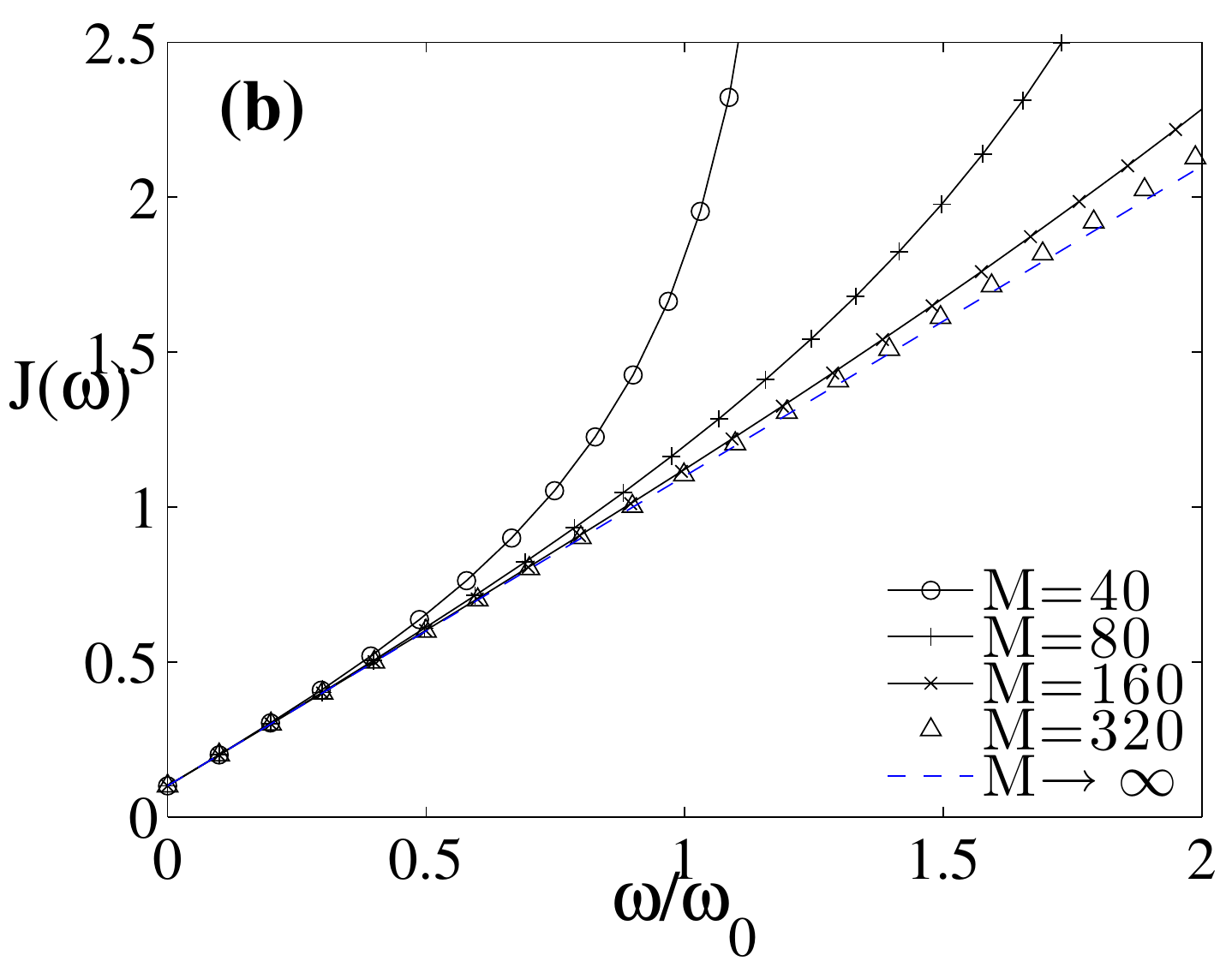}%
  \includegraphics[width=0.33\linewidth]{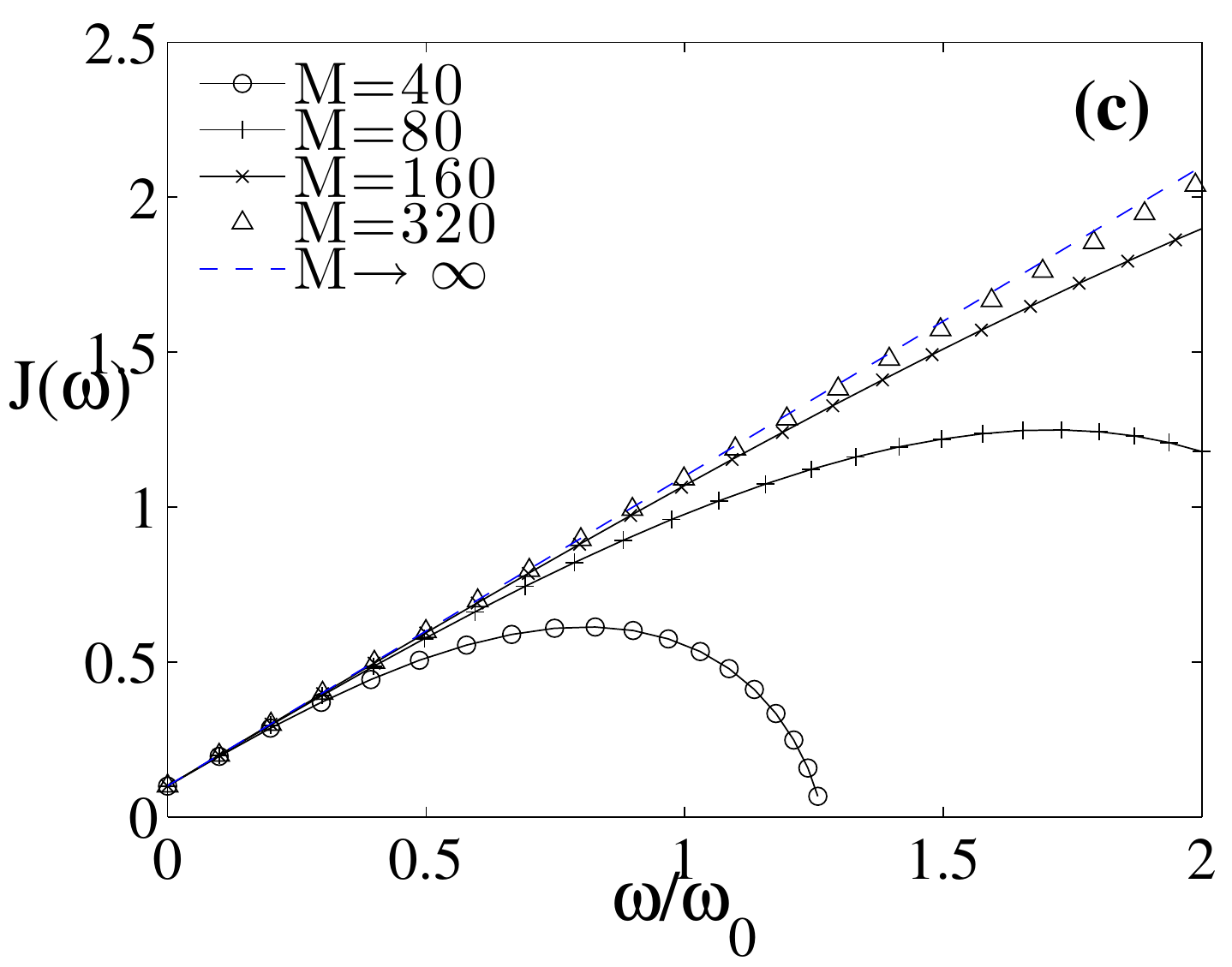}%
  \caption{Spectral properties without the $\mathbf{A}^2$ term ($\Delta=0$), as a function of the normalized frequencies $\nu_k=\omega_k/\omega_0$ and momenta. Mode eigenfrequencies (a) and spectral functions for the capacitive (b) and inductive coupling (c). The simulation assumes a waveguide with $L=10\lambda_0$, where $\lambda_0$ is the wavelength associated to $\omega_0$, and uses $M=40,80,160$ and $320$ modes to extrapolate the dispersion relation. }
  \label{fig:diag-linear}
\end{figure*}

This Letter is structured as follows. We start by introducing a generalised spin-boson model
formally describing a qubit coupled to the open line, including the diamagnetic term, and demonstrating that (i) the spectral function of the model is always Ohmic and (ii) the effective coupling decreases with the strength of the $\mathbf{A}^2$ term. We then specialise the model to describe a particular microscopic setup, namely a transmon capacitively coupled to the transmission line. The microscopic properties of the circuit are related to the parameters in the spin-boson model, allowing us to rigorously discuss the experimental requirements for the observation of the decoupling effect.  We find that the decoupling takes place not only in the DSC regime, but \textit{for all regimes of qubit-line coupling}, opening to the possibility to observe such counter-intuitive effect with present day experiments.

Using the qubit gap $\omega_0$ as unit of energy, the spin-boson model in reads
\begin{equation}
\label{sbmodel}
\ham = \frac{1}{2}\sigma^z + \ham_{\Delta} +  d \sigma^x F,
\end{equation}
The photonic Hamiltonian, $\ham_{\Delta} = \sum_k \nu_k a^\dagger_k a_k + \Delta F^2$, includes the quadratic diamagnetic term weighted by the parameter $\Delta$. The operator $F = \sum_k f_k (a_k + a^\dagger_k)$ depends only on the normal modes $(f_k\propto\sqrt{\nu_k})$, and describes the field that interacts with the qubit through the dipole $d$. 

The strength of the light-matter interaction in the spin-boson model is determined by the spectral function $ J(\nu) = 2\pi\sum_k d^2 |f_k|^2 \delta(\nu - \nu_k)$. When $\Delta=0$ we recover the usual spin-boson model, characterized by the Ohmic spectral function\ \cite{Peropadre13,Leggett87} $J(\nu; \Delta=0)= d^2 \times 2\pi \alpha \nu^1$, where $\alpha$ contains the details of $f_k$. When $\Delta \neq 0$, we can recover again the usual spin-boson model by diagonalising the photonic Hamiltonian $\ham_{\Delta}$, and rewriting the Hamiltonian in terms of its $\Delta$-dependent normal modes,
\begin{equation}
\label{Hren}
  \ham = \frac{1}{2} \sigma^z +
  	\sum_n \left[ {\nu}_n(\Delta) {a}_n^{\dagger} {a}_n 
          + d {f}_n(\Delta)  \sigma^x ({a}_n + {a}_n^\dagger)\right].
\end{equation}
The mode frequencies and couplings now depend on $\Delta$\ \cite{Note2}, but it will be shown that the new spectral function remains Ohmic, albeit with a reduced coefficient ${\alpha}(\Delta)$.

The problem is still computationally underspecified, lacking a precise expression of the field weights $f_k$. We recall a model for a microwave guide as a chain of coupled oscillators, $[q_n,\phi_m]=i\hbar\delta_{nm}$, characterized by the cutoff frequency $\nu_c$
\begin{equation}
\label{lumped}
\ham_\Delta = \frac{1}{2}\sum_{i=0}^{M-1} \left[q_i^2 + \nu_c^2(\phi_i - \phi_{i+1})^2\right].
\end{equation}
For periodic boundary conditions, the lattice is diagonalized by plane waves with quasimomenta $k=2\pi/\len \times \mathbb{Z}$, where $\len = M\delta{x}$ is the actual resonator size and $\delta{x}$ is the spacing of the discretisation. The lattice dispersion is approximately linear around the qubit, $\nu_k=v|k|$, with a large cutoff $\nu_c=v/\delta{x} \gg 1$ and a speed that we use to define a length scale, $v=1$ . This leads to expressions of the field operators $q_j$ and $\phi_j$
\begin{align}
  \label{qp}
  \phi_j & = \sum_{k} \frac{1}{\sqrt{2\nu_k}} \left(\frac{e^{i k x_j}}{\sqrt{M}} a_k + \mathrm{H.c.}\right), \\
  q_j &=  \sum_{k} \sqrt{\frac{\nu_k}{2}}
 \left( \frac{-i e^{i k x_j}}{\sqrt{M}} a_k + \mathrm{H.c.}\right)\;(=\partial_t \phi_j).\nonumber
\end{align}
Note that, while $\phi$ and $q$ are related to the flux and charge operators in circuit-QED, they lack microscopic parameters (inductances, capacitances, etc) that are abstracted into the dipole moment $d$ and quadratic weight $\Delta$.

We contemplate two types of $F$ operators, reproducing the transmon's and charge-qubit's capacitive coupling to the voltage of the transmission line, $F^{(\text{cq})} $, and the flux qubit's inductive coupling to the intensity running through the line, $F^{(\text{fq})}$
\begin{align}
  F^{\text{(cq)}} &\sim \delta{x}^{-1/2} q_x, \\
  F^{\text{(fq)}} &\sim \delta{x}^{-3/2}(\phi_{x+1}-\phi_{x})\nonumber.
\end{align}
These expressions ensure that the spectral function is independent of the discretization at low energies\ \cite{Note1}
\begin{align}
  J^{\text{(cq,fq)}}(\nu;\Delta=0) &= \sum_k 4\pi \frac{\nu_k}{L} d^2 \delta(\nu-\nu_k),
\end{align}
and leads to the expected Ohmic behavior.

We now regard the effect of the $F^2$ term. We recast the resonator Hamiltonian in the matrix form $\ham_\Delta = \frac{1}{2} \mathbf{q}^T \mathcal{\bar{C}}\, \mathbf{q} + {\boldsymbol\phi}^T \mathcal{\bar{L}}\, {\boldsymbol\phi}$, and diagonalize it with a canonical transformation. The new eigenmodes and eigenfrequencies are used to reexpress the $F$ operator and to compute the new spectral function, $J(\nu;\Delta)$. In order to do this, we first fix a length $\len=10 \lambda_0 = 20\pi v/\omega_0$ that ensures a small level spacing $\mathrm{d}\nu = 2\pi v/\len\omega_0 \ll 1$. We then diagonalize the problem for increasingly finer discretizations, $\delta{x}$, doing a finite size scaling to obtain the pairs of frequencies and couplings, $\{\nu_n(\Delta),f_n(\Delta)\}$, in the limit $\delta{x}\to0$. Using these values we compute a function%
\begin{equation}
  D(\nu) = \int_0^\nu d\nu' J(\nu') \simeq 2 \pi \sum_{\nu_n(\Delta) \leq \nu} |d f_n(\Delta)|^2.
\end{equation}
This function is fitted to the appropriate expression and then differentiated to obtain $J(\nu)$. It is quite crucial to work carefully with the interpolation of the function $D(\nu)$, to eliminate finite size and discretization effects that do not contribute in the limit $\delta{x}\to0$. 

\begin{figure}
  \centering
  \includegraphics[width=0.49\linewidth]{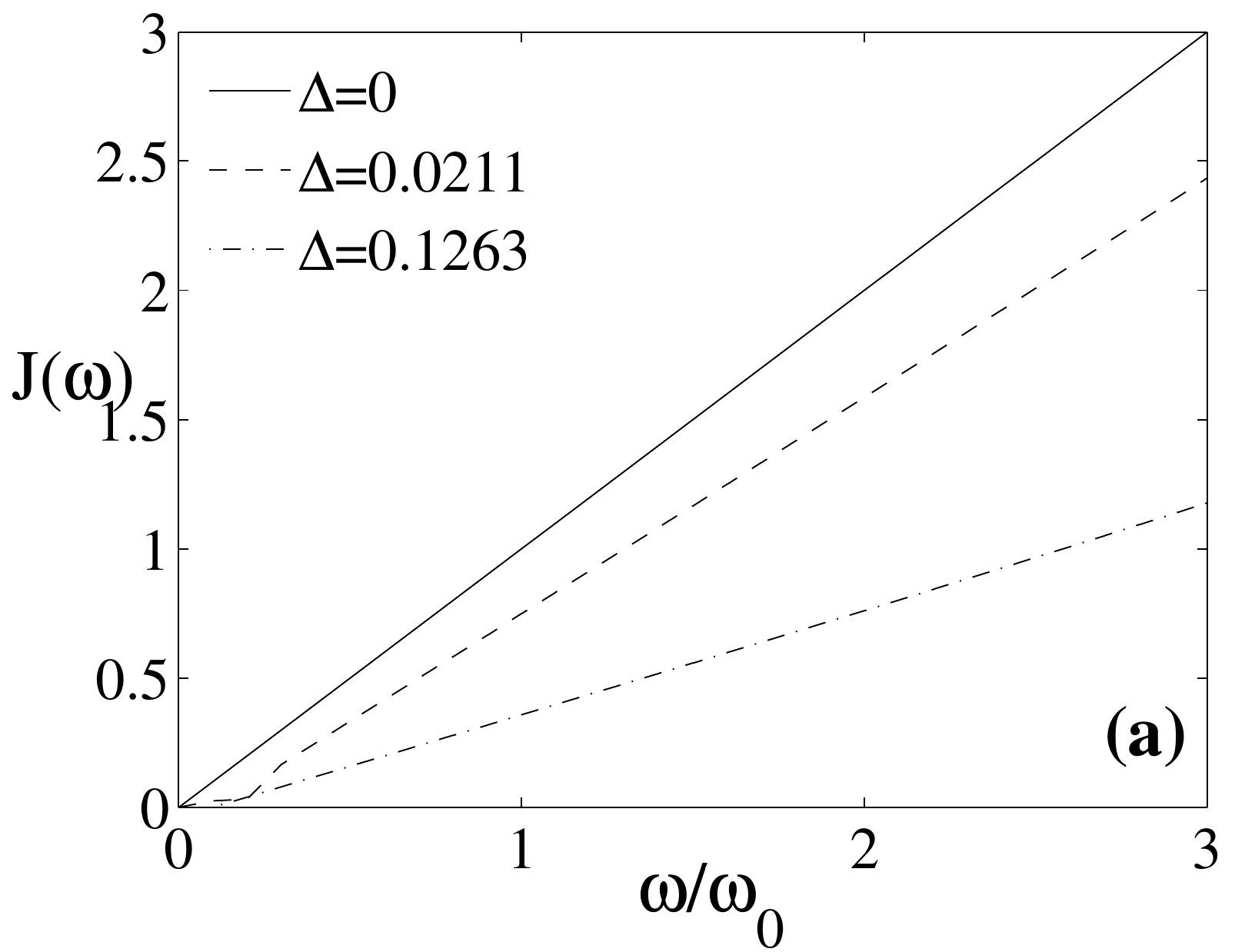}
  \includegraphics[width=0.495\linewidth]{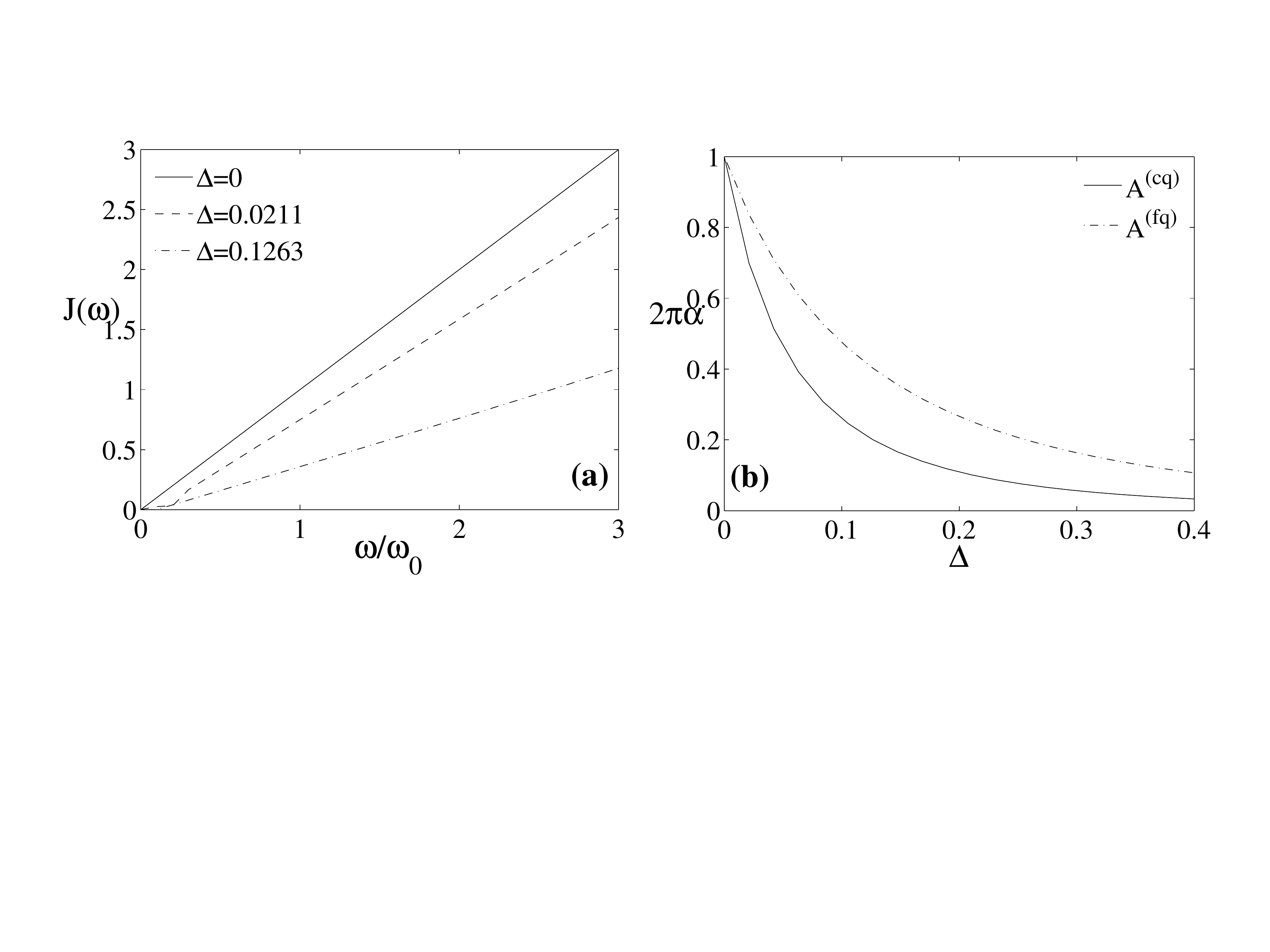}
  \caption{Light-matter decoupling in presence of the $\mathbf{A}^2$ term ($\Delta\neq0$). Panel (a): spectral function for various values of $\Delta$ in a model with $F^{\text{(cq)}}$ coupling. Panel (b): value of the coefficient $\alpha$ in a fit to $J(\nu;\Delta) = 2\pi\alpha(\Delta)\nu^1$ for the couplings $F^{\text{(cq)}}$ and $F^{\text{(fq)}}$. In both sets of simulations $L=10\lambda_0$, and $v=1$. }
  \label{fig:delta-cq}
\end{figure}

In Fig.\ \ref{fig:diag-linear} we exemplify the procedure using the linear case ($\Delta=0$). Note how the frequencies converge to a linear dispersion relation, $\nu_n(0) = 2\pi v/L + \mathcal{O}(\delta x)$, already for 60 modes. The spectral function also comes linear quickly, $J(\nu,0) \sim \alpha \nu^{1} + \mathcal{O}(\delta x)$, allowing us to extrapolate the coefficient $\alpha$ in the continuum limit, $\delta{x}\to0$. Once we learn how to characterize the linear case, we can apply the same procedure to the model with quadratic terms. In Fig.\ \ref{fig:delta-cq}(a) we show the outcome of these simulations for a capacitive coupling, $F^{\text{(cq)}}$. As shown in the plot, the model remains Ohmic, but the slope is changed, decreasing with increasing $\Delta$. In practice this means that the parameter $\alpha$ describing the coupling strength rapidly decreases with $\Delta$. This is observed for both $F^{\text{(cq)}}$ and $F^{\text{(fq)}}$, as shown in Fig.\ \ref{fig:delta-cq}(b). In particular, the capacitive coupling can be fit $2\pi\alpha^{(cq)} = (1 + 6.77 \Delta)^{-2.57}$,
undistinguishable from the actual plot in Fig.\ \ref{fig:delta-cq}b.

\begin{figure}[t]
  \centering
  \includegraphics[width=0.87\linewidth]{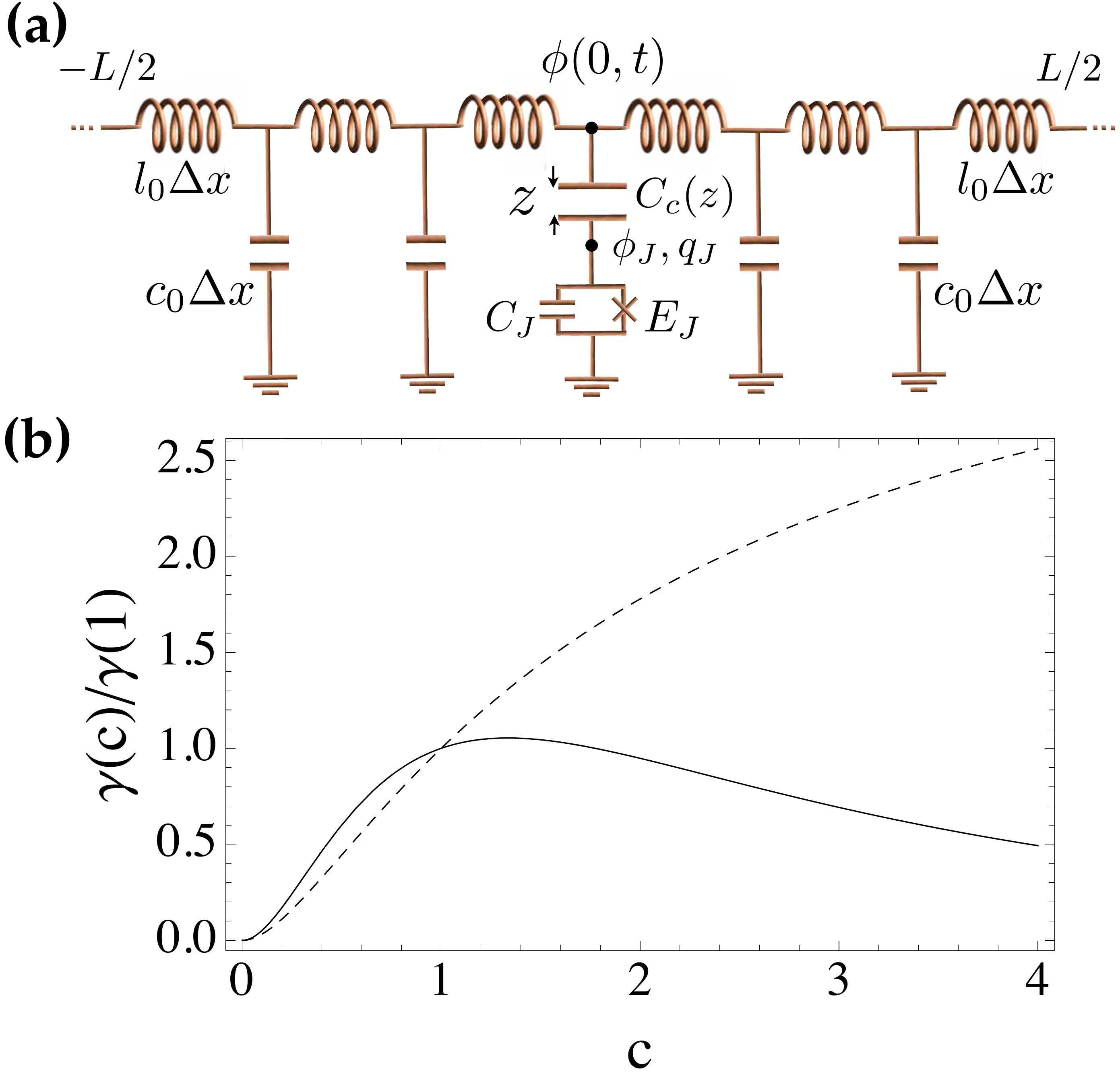}
  \caption{(a) Equivalent circuit of a suspended transmon capacitively coupled to an open transmission line, via a spatial-dependent capacitance $C_c(z)$. (b) Relative value of the spontaneous emission rate as a function of the normalize coupling capacitance, $c=C_c/C_J$, for a model with (solid) and without (dashed) $\mathbf{A}^2$ term. We assume $C_J=25$fF, $Z_0=50\Omega$ and a qubit gap $\omega_0=2\pi\times 7.5$ GHz.}
  \label{fig:decay-rate}
\end{figure}

So far we have been working with a dimensionless rewrite of the spin-boson model, where the microscopic parameters were abstracted into the dipole moment of the qubit $d$ and the weight of the quadratic term $\Delta$. In actual physical systems, both $d$ and $\Delta$ depend on similar physical parameters and, even if $\alpha(\Delta)$ decreases, the product $J(\nu)\propto d^2\alpha(\Delta)$ might not. We therefore now incorporate an actual physical system to see the interplay between both effects. We will study a transmon capacitively coupled to an open transmission line, as depicted in Fig.\ \ref{fig:setup}. The transmon is suspended over the line at a certain height, but the equivalent circuit to this setup, shown in Fig.\ \ref{fig:decay-rate}(a), is similar to the one for an in-plane transmon\ \cite{Peropadre13a}. The circuit Hamiltonian, $H=H_{qb}+H_{int}+H_{\Delta}$, reads
\begin{align}
  H_{qb} &= \frac{q_J^2}{2C_\Sigma} - E_J \cos(2e\phi_J/\hbar),\\
  H_{int} &= \frac{C_c}{C_\Sigma} q_J \partial_t\phi(0,t), \nonumber\\
  H_{\Delta} &= \frac{C_c^2}{2C_\Sigma} \partial_t \phi(0,t)^2 +
  \int_{-L/2}^{L/2} \frac{q(x,t)^2}{2c_0} + \frac{\partial_x\phi(x,t)^2}{2l_0}\mathrm{d}x.\nonumber
\end{align}
The first line provides the qubit eigenenergies, $\frac{1}{2}\hbar\omega_0\sigma^z$, built from the canonical variables of the transmon: the flux $\phi_J$ and the charge $q_J$. The second line is the capacitive coupling between the qubit and the line, $d\sigma^x F$, which is a function of the coupling and total capacitances, $C_c$ and $C_\Sigma=C_c+C_J$. Finally, the third term contains the line Hamiltonian for its charge and flux distributions, $q(x,t)$ and $\phi(x,t)$, renormalized by the quadratic term that arises from the qubit-line coupling. The capacitance and inductance per unit length determine the speed of light $v=(c_0l_0)^{- 1/2}$ and also appear in the particular expression of the coupling operator
\begin{equation} 
  \partial_t\phi(x,t) = \sum_k \sqrt{\frac{\hbar\omega_k}{2c_0}}\left(i\frac{e^{-i(\nu t - kx)}}{\sqrt{L}}a_k+
\mathrm{H.c.}\right).
\end{equation}
Substituting the expression for the field and relating it to the model that we solved numerically before, we obtain
\begin{align}
  \hbar \omega_0 \times d f_k &= \left|\bra{1}{q_J}\ket{0}\right|
  \frac{C_c}{C_\Sigma} \sqrt{\frac{\hbar\omega_k}{2 c_0 L}},\\
  \hbar \omega_0 \times \Delta f_k f_k' &= \frac{C_c^2}{C_\Sigma} \frac{\hbar}{2c_0L} \sqrt{\omega_k\omega_{k'}},\nonumber
\end{align}
where the charge operator is evaluated between two eigenstates of the qubit. We now recall an expression for the capacitive coupling and express it in terms of dimension-full quantities
\begin{equation}
  f_k^{(cq)} = \sqrt{\frac{\omega_k/\omega_0}{2 (L\omega_0/v)}}.
\end{equation}
This leads to the relation
\begin{equation}
  \Delta= \frac{C_c^2}{C_\Sigma}Z_0\omega_0,\; d= \frac{C_c}{C_\Sigma}\times\frac{2e}{\hbar} \bar{n} Z_0^{1/2}
\end{equation}
where $Z_0=\sqrt{l_0/\omega_0}$ is the impedance of the line and $\bar{n}=\left|\bra{1}{q_J}\ket{0}\right|/2e$ is the matrix element of the number operator between two lowest transmon energy levels.

From these expressions it is clear that $J(\omega) =2\pi d^2 \alpha(\Delta) \omega$ cannot have a monotonic behavior with respect to the coupling strength: while $d^2$ grows as $C_c^2$, $\alpha$ decreases with $C_c^2$ and the product of both must saturate or decrease at large couplings. To analyze this behavior we introduce the relative capacitance $c=C_c/C_J$, with which we can express the evolution of the spontaneous emission rate of the qubit onto the line as
\begin{equation}
\frac{\gamma(c)}{\gamma(1)} = \frac{J(\omega; c)}{J(\omega;c=1)} = 4 \frac{c^2}{(1+c)^2} \left[\frac{1+\kappa/2}{1+\kappa\frac{c^2}{1+c}}\right]^{2.57},
\end{equation}
where $\kappa=6.77 C_J Z_0 \omega_0$. This expression has to be compared with the one that we would obtain without $\mathbf{A}^2$ renormalization, which would be
\begin{equation}
\left(\frac{\gamma(c)}{\gamma(1)}\right)_{\Delta=0} = 4 \frac{c^2}{(1+c)^2}.
\end{equation}
Figure\ \ref{fig:decay-rate}(b) shows the behavior of both functions for a transmon qubit with $C_J=25$fF, $Z_0=50\Omega$ and $\omega_0=2\pi\times 7.5$ GHz. Due to the weak anharmonicity, such a qubit \textit{is not in the ultrastrong coupling regime} and is restricted to coupling strengths $g/\omega_0\simeq 5\%$. Nevertheless, even then we show evidence of non-monotonic behavior of the spontaneous emission rate as $\gamma(c)$.

In order to experimentally probe this behavior we would need a qubit with a tunable coupling capacitance. Such a setup already exists: it consists of a mobile transmon that is suspended at a height $z$ on top on the transmission line [cf.\ Fig.\ \ref{fig:setup}], as in the experiment by Houck\ \cite{Shanks13}. By probing different separations between the transmon an a transmission line, and measuring how much energy the qubit deposits onto the transmission line when spontaneously decaying, one should see a similar dependence to the one in Fig.\ \ref{fig:decay-rate}(b), where $c\propto 1/z$. Note that in this analysis we have not considered the coupling to out-of-plane electromagnetic modes as the qubit rises: these decay channels add up to the total emission rate of the qubit, but do not affect the emission into the line and have a negligible contribution of $\mathbf{A}^2$.

Summing up, our study has shown that the $\mathbf{A}^2$ term can decrease the effective light-matter coupling as measured by the Ohmic coupling parameter $\alpha$. In practical setups, the same microscopic parameters that allow increasing the dipolar coupling strength, $d$, also cause a growth of the $\mathbf{A}^2$ term. This has the consequence that the actual coupling strength will eventually saturate and decrease. This non-monotonic behavior occurs for all ranges of the interaction, not only in the ultrastrong or deep coupling regime, as we have shown by analyzing a transmon qubit whose capacitive coupling to a transmission line is modulated. We expect that these results will aid us in better understanding the actual limits in the achievable light-matter interaction strengths, as well as in the design of further superconducting circuits. Moreover, the implementation of an experiment such as the one suggested in this work would represent the first experimental evidence of both the light-matter decoupling effect, as initially predicted for linear systems\ \cite{DeLiberato14} here extended to individual few-level systesm. Moreover, it would also represent the first evidence of the presence of the $\mathbf{A}^2$ term in circuit quantum electrodynamics, and it could open the door to studies of the influence of this term in the superradiant phase transition.

\acknowledgements

This work has been realized with support from the European project PROMISCE, the MINECO Project FIS2012-33022, the EPSRC Project EP/L020335/1 and the CAM Research Consortium QUITEMAD+. Simone De Liberato is Royal Society Research Fellow. B.P. acknowledges support from the STC Center for Integrated Quantum Materials, NSF Grant No. DMR-1231319.

\end{document}